\documentstyle[pra,eqsecnum,aps,amsmath,amssymb]{revtex}
\def\>{\rangle}
\def\<{\langle}
\def\n{\nonumber}
\def\lb{\left[}
\def\rb{\right]}
\def\lk{\left\{}
\def\rk{\right\}}
\def\<{\langle}
\def\>{\rangle}
\def\be#1\ee{\begin{equation}#1\end{equation}}
\def\ba{\begin{eqnarray}}
\def\ea{\end{eqnarray}}

\input{psfig}

\begin{document}
\twocolumn
\widetext

\title{ Entanglement and the Factorization-Approximation}
\author{Jochen Gemmer and G\"unter Mahler}
\address{Institut f\"ur Theoretische Physik,\\Universit\"at Stuttgart, Pfaffenwaldring 57,\\70550 Stuttgart, Germany}
\date{\today}
\maketitle


\begin{abstract}
For a bi-partite quantum system defined in a finite dimensional
Hilbert-space we investigate in what sense entanglement change and
interactions imply each other. For this purpose we introduce an
entanglement-operator, which is then shown to represent a non-conserved
property for any bi-partite system and any type of interaction. This
general relation does not exclude the existence of special initial product
states, for which the entanglement remains small over some period of time,
despite interactions. For this case we derive an approximation to the full
Schr\"odinger-equation, which allows the treatment of the composite systems
in terms of product states. The induced error is  estimated. In this factorization-approximation one
subsystem appears as an effective potential for the other. A pertinent
example is the Jaynes-Cummings model, which then reduces to the
semi-classical rotating wave approximation. 
\end{abstract}

\narrowtext
\section{Introduction}
During the last decades entanglement has been investigated under various
aspects. The famous EPR-paradox, for example, has led to a discussion of the most basic
principles of quantum physics \cite{EIN35}. The Gedanken experiment based
on "Schr\"odinger's Cat" may be seen as an attempt to challenge the
consistency of quantum
mechanics: It has been argued that this situation could only be understood
by allowing for entanglement between the atom and the cat which,
on the other hand, should be considered a classical object
\cite{OTT98}. But, by definition, a classical object cannot become
entangled with any other system.\\
Since it has been shown, that quantum algorithms have the potential to
outperform corresponding classical computing \cite{DEU85} \cite{GRO97}
\cite{SHO94}, considerable efforts have been made to implement gates like
the so-called quantum controlled NOT-gate (QCNOT). Performing a  QCNOT
generically results in preparing an entangled state. Meanwhile various
experimental schemes to prepare entangled states have been developed
\cite{CHU98}, \cite{TUR98}.\\
In all these approaches entanglement has been in the very center of interest. The question was always either how to interprete the state of two systems being entangled, or how to deliberately produce entanglement and detect it, once it has been produced.\\
Rather neglected seems to have been the question of entanglement as an unavoidable "waste product" of
quantum mechanical dynamics. Little attention has been paid to the fact
that it cannot be taken for granted that any two interacting systems will
remain in a product state, even if they have been in one in the begining
\cite{ESP90} \cite{LUI98}. This means that there is always the possibility
for them to entangle. And if they are entangled, it is impossible to assign
two separate wavefunctions to the subsystems. Nevertheless this is typically done in standard "textbook level" quantum mechanics: The particle in a box, e.g., is always described by a wavefunction although it definitely interacts with the box that neccesarily consists of a many particle-quantum-system itself and therefore could become entangled with it. There is no discussion of the electron going through the double slit being possibly entangled with the material defining the slit itself.\\
But since these approximations typically lead to excellent results, it should be possible to point out why. In which situations is it reasonable to neglect entanglement and treat whole complicated systems as effective potentials for another quantum-system?\\
Apart from the rather academic desire to undestand the basis of this
"classical limit", there is also a good practical reason to adress such questions.\\
An important prerequisit of all quantum computer designs suggested so far is the possibility of so called local unitary
transformations. These should be performed selectively on each effective spin
(q-bit) through potentials that are supposed to be controllable in time
\cite{KAN98}. But again, in reality, those potentials can only be
implemented by means of other complicated quantum-systems that could
possibly entangle with those spins: this would inevitabely lead to
decoherence. But quantum computers need to be coherent. In that sense the
problem of entanglement through interaction (as required by external
control) could even challenge the implementation of any real quantum
computer.\\
Our paper is organized as follows: We first specify a theorem
relating some purity measure $P$ (as an entanglement test) to inter-subsystem interactions (Sect. \ref{theo}). For the proof of this theorem
(Sect. \ref{protheo}) we proceed as follows:\\
Starting from the Von-Neumann-equation, which describes the dynamics of the
density operator, we proceed by inserting an expansion of the density operator into this equation. The result is an equation only in terms of the expansion coefficients that has exactly the form of the Schr\"odinger-Equation and will therefore be called "quasi-Schr\"odinger-equation". It is now possible to define a linear operator in the space of those coefficients which has an expectation value equal to $P$, and will therefore be called "purity operator". Since the dynamics of those coefficients are controlled by an equation that is formally identical with the Schr\"odinger-equation (including a ``quasi-Hamiltonian''), it is possible to reduce the question of $P$ being conserved or not, to the problem whether the commutator of the purity operator and quasi-Hamiltonian will vanish or not.\\ 
Thus, the mathematical scheme used here is essentially the same as used in standard quantum mechanics to identify conserved quantities. Only the space of the state vector and the interpretation of the considered quantities, are different.\\
The last step will be to show, that the above commutator becomes nonzero
whenever the full Hamiltonian involves any kind of interaction.\\
However, even in the presence of interactions the system may remain
``almost'' unentangled. In Sect. \ref{facta} we use our quasi-Schr\"odinger formulation to derive the
factorization-approximation with its effective potentials for this case. In
Sect. \ref{erres} the induced error is estimated to lowest order.
In Sect. \ref{leckm} we apply the results to the Jaynes-Cummings-model.

\section{theorems}
\label{theo}
There is a still ongoing debate on entanglement measures \cite{VED97}. A lot of
propositions have been made, but it seems still rather difficult to
introduce a general entanglement measure that satisfies all conditions that
have been imposed on such a measure and, at the same time, is applicable
for any  number of subsystems and any case (pure and mixed states of the
whole system). And it seems even more difficult to construct a measure in such a way that it could actually be calculated (or measured!) for reasonably complicated situations.\\ 
Fortunately, it is possible to introduce a simple measure under specific
conditions: If the state of the whole system is a pure state, and the full
system is being regarded as divided into two subsystems, a convenient
entanglement measure is $1-P$, where
\be
P=\mbox{Tr}_I\lk (\rho^I)^2\rk=\mbox{Tr}_{II}\lk (\rho^{II})^2\rk
\ee
(here $\rho^I,\rho^{II}$ are the reduced density matrices of the
corresponding subsystems)\\
Entanglement between to subsystems originating from unitary quantum
evolution, can only result from interactions \footnote{Non-unitary
  transformations can do without direct interactions: This phenomenon has
  become known as entanglement swapping \cite{JIA98}.}. If two systems do not interact they can be treated without even taking the other one into account. So, if they are both in pure states at the begining, which means they are in a product state with respect to the whole system, they will remain so forever under these conditions.\\
One may ask now whether two systems that interact might remain entanglement-free, depending for example on the kind of systems that interact, or on the kind of interaction that is considered.\\
Concerning this question we are aware of only rather vague statements in
the literature. A typical formulation due to d'Espagnat reads:\\
{\bf Theorem A}:\\
"In general it is impossible to describe systems that interacted in the
past by separate wavefunctions" \cite{ESP90}.\\
But does this always have to be the case? To address this problem we will
prove the following theorem for finite discrete Hilbert-spaces:\\
{\bf Theorem B}:\\
"There exists no interaction what so ever between arbitrary systems, such
that the entanglement measure ($1-P$) remains conserved".\\
This theorem does not imply that there cannot be initial states, starting
from which the sytem might remain in a product state, though it can be shown
that those states, if they exist at all, only play a negligible role in
typical larger systems. But it definitely means, that there must be initial states that lead to entanglement, even between the particle and the box-system, or between the electron and the slit-system.\\ 
Further consequences of theorem B can most conveniently be assessed from an
approximation-scheme that is valid as long as the systems remain
approximately unentangled, as  will be shown in Sect. \ref{facta}, \ref{erres} and Sect. \ref{leckm}.\\

\section{Proof of Theorem B}
\label{protheo}

\subsection{Basis operators}
The operators into which the density matrix will be expanded here, are products of the generators of the respective $SU(n)$ groups,
where one set of generators corresponds to one subsystem \cite{MAH98}.\\
The basis operators for the different subsystems are defined in the following way:
\be
\label{def}
\hat{Q}_{i_\nu}: = \left\{ \begin{array}{ccc} \frac{1}{\sqrt{n_\nu}}\hat{1}^\nu&:&i=0\\ \frac{1}{\sqrt{2}}\hat{\lambda}^\nu_i&:&i\neq0 \end{array} \right.
\ee
where, $\nu$ denotes the index of the subsystem, $n_\nu$ the number of
levels of subsystem $\nu$ (dimension) and $\hat{\lambda}^\nu_i$ the generator of the $SU(n_\nu)$ group, $i=1,2\ldots n_{\nu}^2-1$.\\
The basis operators of the full system are defined as dyadic products of the basis operators of the subsystems,
\be
\label{defprod}
\hat{Q}_{\vec{i}}:=\prod^N_{\nu=0}\! \otimes  \,\hat{Q}_{i_\nu}
\ee
where $N$ is the number of subsystems.\\
Thus any basis operator is defined by a sequence of $N$ indices $i_{\nu}$
(abbreviated as $\vec{i}$), each index specifying, which generator $i$ should be
applied to the corresponding subsystem $\nu$. The operators constructed according to these rules form a complete and orthonormal set in the following sense:
\be
\label{ort}
\mbox{Tr}\{\hat{Q}_{\vec{i}}
\hat{Q}_{\vec{j}}\}=\delta_{\vec{i}\vec{j}}:=\prod^N_{\nu=1}\delta_{i_{\nu}j_{\nu}}\quad \hat{A}=\sum_{\vec{i}}\hat{Q}_{\vec{i}}\mbox{Tr}\{\hat{Q}_{\vec{i}}\hat{A}\}
\ee
where $\hat{A}$ is an arbitrary operator.\\
Representing the density matrix in the case of only two subsystems as
\be
\label{expdens}
\hat{\rho}=\sum_{i,j}q_{ij}\hat{Q }_i \otimes \hat{Q }_j
\ee
where the index $i$ corresponds to
subsystem I and the index $j$ to subsystem II, we find
\be
\label{reddens}
\hat{\rho}^I=\sqrt{n_{II}}\sum_i q_{i0}\hat{Q }_i\quad P=n_{II}\sum_iq_{i0}^2.
\ee
The objects we are going to examine aer thus specified in terms of their
expansion coefficients.

\subsection{Quasi-Schr\"odinger-Equation}
The Von-Neumann-equation reads:
\be
i\hbar\frac{d}{dt}\hat{\rho}=\lb\hat{H},\hat{\rho}\rb.
\ee
Inserting the expansion (\ref{expdens}) yields:
\be
i\hbar\sum_{\vec{i}}\frac{d}{dt}q_{\vec{i}}\hat{Q}_{\vec{i}}=\sum_{\vec{j}}\lb\hat{H},\hat{Q}_{\vec{j}}\rb q_{\vec{j}}
\ee
After multiplying by $\hat{Q}_{\vec{m}}$, taking the trace and applying
some trace theorems, we get:
\be
\label{quasimat}
i\hbar\frac{d}{dt}q_{\vec{m}}=\sum_{\vec{j}}{\mathcal H}_{\vec{m}\vec{j}}
q_{\vec{j}}\quad {\mathcal H}_{\vec{m}\vec{j}}:=\mbox{Tr}\lk\hat{H}\lb\hat{Q}_{\vec{j}},\hat{Q}_{\vec{m}}\rb\rk
\ee
This equation (hereafter called quasi-Schr\"odinger-equation) has evidently the
Schr\"odinger-form. The hermiticity of ${\mathcal{H}}$ is easily shown by examining the corresponding matrix elements,
\ba
&&{{\mathcal H}}^*_{\vec{m}\vec{j}}= \mbox{Tr}\lk\hat{H}\lb\hat{Q}_{\vec{j}},\hat{Q}_{\vec{m}}\rb\rk^*=\mbox{Tr}\lk\left(\hat{H}\lb\hat{Q}_{\vec{j}},\hat{Q}_{\vec{m}}\rb\right)^{\dagger}\rk\\
&&=\mbox{Tr}\lk\hat{H}\lb\hat{Q}_{\vec{m}},\hat{Q}_{\vec{j}}\rb\rk={\mathcal
  H}_{\vec{j}\vec{m}}\n
\ea
We can even define a formal bracket notation: For this purpose we re-arrange the
multiple indices $\vec{j}$ as a simple index $s$ and introduce a set of real orthogonal basis vectors
\be
|s\>=|s\>^*
\ee
with
\be
\<s|s'\>=\delta_{ss'} \qquad \hat{1}=\sum_s|s\>\<s|
\ee
such that
\be
\label{matel}
{\mathcal H}_{ss'}=\<s|{\mathcal H}|s'\> \qquad  q_s=\<s|q\>
\ee
Inserting (\ref{matel}) into (\ref{quasimat}) yields:
\be
i\hbar\frac{d}{dt}|q\>={\mathcal{H}}|q\>
\ee
with the infinitesimal solution
\be
\label{evoeq}
|q(dt)\>=\left(1+\frac{1}{i\hbar}{\mathcal{H}}dt\right)|q(0)\>
\ee
On the basis of this formal equivalence it is now possible to investigate the
conservation of some quantity ${\mathcal A}$ in the space of the vectors $|q\>$ ($\<q|{\mathcal A}|q\>=\sum_{s,s'}q_s{\mathcal
  A}_{ss'}q_{s'}$), by evaluating the commutator of ${\mathcal A}$ with ${\mathcal{H}}$.

\subsection{Purity Operator}
We first note that our purity measure, $P$, can be given the  mathematical
form of an expectation value:
\be
\<q|{\mathcal P}|q\>=P
\ee
To find ${\mathcal{P}}$, we go back to the explicit multi-index notation
(for two subsystems),
\be
P=\sum_{s,s'}q_s{\mathcal
  P}_{ss'}q_{s'}=\sum_{i,j,u,v}q_{ij}{\mathcal P}_{ijuv}q_{uv}.
\ee
If we require (see (\ref{reddens}))
\be
\<q|{\mathcal P}|q\>=n_{II}\sum_iq_{i0}^2
\ee
it follows that
\be
P_{ijuv}=n_{II}\delta_{iu}\delta_{jv}\delta_{0j}
\ee
i. e., up to a normalization factor ${\mathcal{P}}$ is a projector, projecting
out the components of $|q\>$ that refer locally to subsystem I, those
components that would read $q_{i0}$ in multi-index notation for a
bi-partite system. Using the bracket notation ${\mathcal{P}}$ reads:
\be
{\mathcal{P}}=n_{II}\sum_w|w^I\>\<w^I|
\ee
Adding the complementary projector that projects out all the other components, that
is all those that do not refer locally to subsystem I, we can write the
unity operator as:
\be
\hat{1}=\sum_w|w^I\>\<w^I|+\sum_k|k^R\>\<k^R|
\ee
Using this representation ${\mathcal{H}}$, which controls the complete
dynamics, can be split up into:
\be
\label{hsplit}
{\mathcal H}={\mathcal L}^I+{\mathcal R}+{\mathcal W}
\ee
where
\ba
{\mathcal L}^I&:=&\sum_{l,m}{\mathcal L}^I_{lm}|l^I\>\<m^I|\n\\
{\mathcal R}&:=&\sum_{i,j}{\mathcal R}_{ij}|i^R\>\<j^R|\n\\
{\mathcal W}&:=&\sum_{l,j}\left({\mathcal W}_{lj}|l^I\>\<j^R|+{\mathcal W}_{jl}|j^R\>\<l^I|\right)\n.
\ea
One easily convinces oneself, that
\be
\lb{\mathcal{P}},{\mathcal{L}}^I\rb=0\quad \mbox{and} \quad \lb{\mathcal{P}},{\mathcal{R}}\rb=0.
\ee
Finally, the commutator $\lb{\mathcal{P}},{\mathcal{W}}\rb$ reads:
\be
\label{com}
\lb{\mathcal{P}},{\mathcal{W}}\rb=n_{II}^2\sum_{w,j}\left({\mathcal W}_{wj}|w^I\>\<j^R|-{\mathcal W}_{jw}|j^R\>\<w^I|\right).
\ee
We now convince ourselves that not all matrix-elements ${\mathcal W}_{wk}$ can be
identically zero in the presence of interactions. For this purpose we define
``interaction'' operationally: Two subsystems are said to interact, if the
dynamics of one subsystem, at least for some initial state (but generically
for any initial state) depend on the state of the other subsystem.\\
The state of subsystem I at time $dt$ is completely discribed by the
projection
\be
|q^I(dt)\>:=\sum_w|w^I\>\<w^I|q(dt)\>.
\ee
Substituting $|q(dt)\>$ according to the evolution equation (\ref{evoeq}) with
${\mathcal{H}}$ given by (\ref{hsplit}) we obtain
\ba
|q^I(dt)\>&=&|q^I(0)\>\\
+\frac{1}{i\hbar}dt&&\hspace{-0.5cm}\left(\sum_{w,m}L^I_{wm}|w^I\>\<m^I|q(0)\>+\sum_{w,j}W_{wj}|w^I\>\<j^R|q(0)\>\right)\n
\ea
As only the components $\<j^R|q\>$ carry information about subsystem II not
all $W$-matrix-elements can be zero for interacting subsystems.\\
Since the two terms that are summed over in (\ref{com}) obviously belong to different (off diagonal) parts of ${\mathcal{H}}$, they cannot cancel each other. Thus we conclude,
\be
\lb{\mathcal{P}},{\mathcal{H}}\rb\neq0,
\ee 
for any sort of interaction, which completes the proof of theorem B. 

\section{Factorization-approximation and effective potentials}
\label{facta}
As we have shown, there is, a priori, no reason to assume
that two interacting systems remain in pure states. Nevertheless, as we
will argue now, it is a reasonable approximation to treat one subsystem as if
it was in a pure state, and the other one as an effective potential for the
former, as long as $1-P$ remains small over the period of time under investigation.

\subsection{Factorization-approximation}
We start with the quasi-Schr\"odinger-equation (\ref{quasimat}) for two subsystems,
\be
i\hbar\frac{d}{dt}q_{ij}=-\sum_{l,m}\mbox{Tr}\lk\hat{H}\lb\hat{Q}_{ij},\hat{Q}_{lm}\rb\rk q_{lm}
\ee
and consider only those equations which refer to the two subsystems
locally, not to the correlations. We thus need to study:
\ba
\label{qualoc}
I:\quad i\hbar\sqrt{n_{II}}\frac{d}{dt}q_{i0}&=&-\sum_{l,m}\mbox{Tr}\lk\hat{H}\lb\frac{\hat{\lambda}_i}{\sqrt{2}},\hat{Q}_{lm}\rb\rk q_{lm}\\
II:\quad i\hbar\sqrt{n_I}\frac{d}{dt}q_{0j}&=&-\sum_{l,m}\mbox{Tr}\lk\hat{H}\lb\frac{\hat{\lambda}_j}{\sqrt{2}},\hat{Q}_{lm}\rb\rk q_{lm}\n
\ea
i.e. we have to examine the relation between ``local'' ($q_{i0},q_{0j}$) and
``global'' ($q_{ij}$) coefficients. We define a tensor $M$ in the following way:
\be
M_{ij}:=\<\hat{Q}_i\otimes\hat{Q}_j\>-\<\hat{Q}_i\>\<\hat{Q}_j\>
\ee
or, according to (\ref{expdens}) and (\ref{ort}):
\be
\label{mdef}
M_{ij}=q_{ij}-\sqrt{n_{I}n_{II}}q_{i0}q_{0j}
\ee
Obviously, for a product state all $M_{ij}$'s have to vanish. If the two
subsystems get correlated (entangled), the $M_{ij}$'s will take on nonzero
values, so that 
\be
\beta:=\sum_{i,j}M_{ij}^2
\ee
can be considered as an alternative entanglement measure. It is indeed possible to derive a relation between $\beta$ and $P$ \cite{SCH95}:
\be
\label{betap}
\beta<1-P^2
\ee
If we now solve (\ref{mdef}) for $q_{ij}$ and insert the result into
(\ref{qualoc}) we get for subsystem I:
\ba
\label{factap}
&&i\hbar\sqrt{n_{II}}\frac{d}{dt}q_{i0}=\\
&&-\sum_{l,m}\mbox{Tr}\lk\hat{H}\lb\frac{\hat{\lambda}_i}{\sqrt{2}},\hat{Q}_{lm}\rb\rk \left(\sqrt{n_In_{II}}q_{l0}q_{0m}+M_{lm}\right)\n
\ea
Using (\ref{reddens}) and (\ref{betap}) we find
\be 
n_In_{II}\sum_{l,m}q_{l0}^2q_{0m}^2=P^2  \qquad \sum_{l,m}M_{lm} ^2<1-P^2,
\ee
so that it seems reasonable to
neglect  the $M_{lm}$-term and only keep
the $\sqrt{n_In_{II}}q_{l0}q_{0m}$-term in (\ref{factap}), as long as $P$ stays
close to 1. The error that occurs if this approximation is used over a
period of time instead of the true Schr\"odinger-equation, will be
estimated later.\\
With the above approximation we get after performing the sum over $m$:
\be
i\hbar\sqrt{n_{II}}\frac{d}{dt}q_{i0}=-\sum_l\mbox{Tr}\lk\hat{H}\lb\frac{\hat{\lambda}_i}{\sqrt{2}},\sqrt{\frac{n_{II}}{2}}\hat{\lambda}_l\rb\otimes\hat{\rho}^{II}\rk q_{l0}
\ee
The latter is a quasi-Schr\"odinger-equation for subsystem I with a
quasi-Hamiltonian depending on the momentary local state of subsystem
II. It is easy to check that the quasi-Hamiltonian of this equation, although depending on time, will always be hermitian. This means that in this approximation  $\sqrt{n_{II}}\sum_iq_{i0}^2=P$ is a conserved quantity. Both subsystems will remain in pure states and unentangled if the initial state was a product state.\\
Performing the sum over $l$ and taking the partial trace with respect to subsystem II yields:
\be
i\hbar\sqrt{n_{II}}\frac{d}{dt}q_{i0}=-\mbox{Tr}_I\lk\mbox{Tr}_{II}\lk\hat{H}\hat{\rho}^{II}\rk\lb\frac{\hat{\lambda}_i}{\sqrt{2}},\hat{\rho}^I\rb\rk
\ee
Since $\hat{\rho}_1$ and $\hat{\rho}_2$ always represent pure states in
this case, we can use separate wavefunctions to re-express lokal
expectation values, such as:
\be
\mbox{Tr}_{II}\lk\hat{H}\hat{\rho}^{II}\rk=\<\psi^{II}|\hat{H}|\psi^{II}\>.
\ee
Attention should be paid to the fact that this expectation value is still an operator with respect to subsystem I, if $\hat{H}$ contains interactions.\\
Following the standard trace theorems, we get:
\be
i\hbar\sqrt{n_{II}}\frac{d}{dt}q_{i0}=\mbox{Tr}_I\lk\frac{\hat{\lambda_i}}{\sqrt{2}}\lb\<\psi^{II}|\hat{H}|\psi^{II}\>,\hat{\rho}^I\rb\rk.
\ee
Multiplying the $i$-th equation by $\frac{\hat{\lambda_i}}{\sqrt{2}}$,
summing over $i$, exploiting the completeness of the
$\frac{\hat{\lambda_i}}{\sqrt{2}}$'s and using (\ref{reddens}) leads to:
\be
i\hbar\frac{d}{dt}\hat{\rho}_I=\lb\<\psi^{II}|\hat{H}|\psi^{II}\>,\hat{\rho}^I\rb.
\ee
This equation is of the same form as the Von-Neumann-equation. Since the
reduced density operators $\hat{\rho}_I,\hat{\rho}_{II}$ represent pure states
we can change to the
corresponding Schr\"odinger-equation, without loss of generality:
\be
i\hbar\frac{\partial}{\partial t}|\psi^I\>=\<\psi^{II}|\hat{H}|\psi^{II}\>|\psi^I\>
\ee
After splitting the Hamiltonian $\hat{H}$ into local ($\hat{L}_I, \hat{L}_{II}$) and interaction ($\hat{W}$) parts and adding the corresponding equation of subsystem II, we find the complete system of equations:
\ba
i\hbar\frac{\partial}{\partial t}|\psi^I\>&=&\left(\hat{L}_I+\<\psi^{II}|\hat{L}_{II}|\psi^{II}\>+\<\psi^{II}|\hat{W}|\psi^{II}\>\right)|\psi^{I}\>\\
i\hbar\frac{\partial}{\partial t}|\psi^{II}\>&=&\left(\hat{L}_{II}+\<\psi^I|\hat{L}_I|\psi^I\>+\<\psi^I|\hat{W}|\psi^I\>\right)|\psi^{II}\>\n
\ea
\subsection{Gauge}
Since the overall phases of both subsystems are arbitrary and can be chosen
freely, a method, which is completely analogous to the Lorentz gauge in
classical electrodynamics, can be used to simplify these equations even further.\\
Applying the substitutions
\be
e^{i\alpha_I(t)}|\phi^I\>:=|\psi^I\> \qquad e^{i\alpha_{II}(t)}|\phi^{II}\>:=|\psi^{II}\>
\ee
and choosing the phases as
\ba
\<\psi^{II}|\hat{L}_{II}|\psi^{II}\>+\hbar\dot{\alpha}_I(t)&=&0\\  
\<\psi^I|\hat{L}_I|\psi^I\>+\hbar\dot{\alpha}_{II}(t)&=&0\n
\ea
transforms the system of equations into:
\ba
\label{fakt}
i\hbar\frac{\partial}{\partial t}|\phi^I\>&=&\left(\hat{L}_I+\<\phi^{II}|\hat{W}|\phi^{II}\>\right)|\phi^I\>\\
i\hbar\frac{\partial}{\partial t}|\phi^{II}\>&=&\left(\hat{L}_{II}+\<\phi^I|\hat{W}|\phi^I\>\right)|\phi^{II}\>.\n
\ea
This is a coupled set of nonlinear first-order differential equations
which, although it does not necessarily create unitary dymamics, keeps the
absolute values of the wavefunctions of both subsystems fixed. Instead of
the $(n_I\cdot n_{II})$ dimensions of the exact treatment, this approximation
has only $(n_I+n_{II})$ dimensions (number of equations).\\
Each of these equations can be considered as an "ordinary"
Schr\"odinger-equation of one system, in which the influence of the other
one appears as an effective potential. This approach may be termed
"quasi-classical" with respect to interactions. It underlies, e.g. the
potential for the particle in a box.\\
Similar equations are well-known in the theory of many particle systems
(for example the Hartree-equation \cite{HAR28}). But in those cases they
are basically justified by their success. We are now going to derive a
criterion for their applicability. 

\section{Error estimation}
\label{erres}
We will now examine the deviation of the solution in
factorization-approximation $|\phi(t)\>$ from the solution of the full
Schr\"odinger-equation $|\psi(t)\>$, under the condition that
$|\phi(0)\>=|\psi(0)\>$ . We will expand this deviation in terms of an
effective interaction which will allow us to connect the resulting error with the purity $P$.\\
The Schr\"odinger-equation reads:
\be
i\hbar\frac{\partial}{\partial t}|\psi\>=\left(\hat{L}_I+\hat{L}_{II}+\hat{W}\right)|\psi\>
\ee
With $|\phi\>:=|\phi^{I}\>\otimes|\phi^{II}\>$ the factorization-approximation reads:
\be
i\hbar\frac{\partial}{\partial t}|\phi\>=\left(\hat{L}_I+\hat{L}_{II}\<\phi^{I}|\hat{W}|\phi^{I}\>+\<\phi^{II}|\hat{W}|\phi^{II}\>+\alpha\right)|\phi\>
\ee
where $\alpha$ denotes an insignificant overall phase, and 
\be
|\phi\>:=|\phi^{I}\>\otimes|\phi^{II}\>
\ee
Introducing the deviation $|\theta\>$ as 
\be
|\psi\>=:|\theta\>+|\phi\>,
\ee
and the abbreviation
\be
\hat{V}:=\hat{W}-\<\phi^{I}|\hat{W}|\phi^{I}\>-\<\phi^{II}|\hat{W}|\phi^{II}\>-\alpha
\ee
we can write a differential equation for this deviation $|\theta\>$ just in terms of
$|\phi\>$ and $|\theta\>$ as 
\be
\label{bas}
\frac{\partial}{\partial t}|\theta\>=\frac{1}{i\hbar}\left(\hat{H}|\theta\>+\hat{V}|\phi\>\right)
\ee
Defining
\be
|\theta'\>:=e^{\frac{i\hat{H}t}{\hbar}}|\theta\>
\ee
we can rewrite (\ref{bas}) as
\be
\label{thepr}
\frac{\partial}{\partial
  t}|\theta'\>=\frac{1}{i\hbar}e^{\frac{i\hat{H}t}{\hbar}}\hat{V}|\phi\>
\ee
The exponential may be written as
\be
\label{expex}
e^{\frac{i\hat{H}t}{\hbar}}=\sum_n|\psi_n(0)\>\<\psi_n(t)|=\sum_n|\phi_n(0)\>\left(\<\phi_n(t)|+\<\theta_n(t)|\right)
\ee
Here the $|\phi_n(0)\>=|\psi_n(0)\>$ are chosen to form a complete
orthonormal set. The $|\phi_n(t)\>$ as solutions in factorization
approximation form a complete orthonormal set
at all times if they did so at $t=0$, so they will be used as a
basis. Since they all obey (\ref{thepr}) we get, using (\ref{expex}):
\be
|\theta_i'(t)\>=\frac{1}{i\hbar}\sum_n|\phi_n(0)\>\int_0^t\left(\<\phi_n(t')|+\<\theta_n(t')|\right)\hat{V}|\phi_i(t')\>dt'
\ee
Transforming
back to the unprimed quantities, we find
\ba
&&|\theta_i(t)\>=\\
&&\frac{1}{i\hbar}\sum_n\left(|\phi_n(t)\>+|\theta_n(t)\>\right)\int_0^t\left(\<\phi_n(t')|+\<\theta_n(t')|\right)\hat{V}|\phi_i(t')\>dt'\n
\ea
By iterating this equation we can produce an expansion of
$|\theta_i(t)\>$ in terms of time integrals over matrix elments of the
effective interaction $\hat{V}$. As long as those matrix elments remain small, a
truncation scheme can be applied, which gives to first order:
\be
\label{tru}
|\theta_i(t)\>=\frac{1}{i\hbar}\sum_n|\phi_n(t)\>\int_0^t\<\phi_n(t')|\hat{V}|\phi_i(t')\>dt'
\ee
We return now to the double-index notation
\be
|\phi_{ij}\>:=|\phi_i^I\>\otimes |\phi_j^{II}\>,\quad \<\phi_i^I|\phi_{i'}^I\>=\delta_{i'i},\quad \<\phi_j^{II}|\phi_{j'}^{II}\>=\delta_{j'j}
\ee
and take $|\psi_{00}\>(|\phi_{00}\>)$ as the solution(approximation) under consideration.
We use the $|\phi_{ij}\>$'s as a basis to write the full solution
\be
\label{ent}
|\psi_{00}\>=\sqrt{1-\sum_{i,j}|\theta_{ij}|^2}|\phi_{00}\>+\sum_{i,j}\theta_{ij}|\phi_{ij}\>,\quad
\theta_{00}:=0
\ee
Its overlap of this solution with the solution in factorization-approximation (``fidelity'') is exactly given by
\be
|\<\psi_{00}|\phi_{00}\>|^2=1-\sum_{i,j}|\theta_{ij}|^2
\ee
On the other hand, the squareroot of the purity of either subsystem of the 
state $|\psi_{00}\>$ to first order in $|\theta_{ij}|^2$ is given by
\be
\sqrt{P}=1-\sum_{i\neq 0,j\neq 0}|\theta_{ij}|^2
\ee
Obviously the two quantities are the same except for the $\theta_{i0}$'s and
the $\theta_{0j}$'s. Computing the $\theta_{ij}$'s from equation
(\ref{tru}) yields:
\ba
&&\theta_{ij}=\frac{1}{i\hbar}\int_0^t\left(\<\phi_{ij}|\hat{W}|\phi_{00}\>-\<\phi_{0j}|\hat{W}|\phi_{00}\>\delta_{i0}-\right.\\
&&\left.\<\phi_{i0}|\hat{W}|\phi_{00}\>\delta_{0j}+\alpha
  \delta_{i0}\delta_{0j}\right)dt'\n
\ea
Choosing $\alpha:=\<\phi_{00}|\hat{W}|\phi_{00}\>$ to satisfy the
definition $\theta_{00}:=0$
from (\ref{ent}) we get
\be
\theta_{i0}(t)=0 \quad \theta_{0j}(t)=0 
\ee
so that
\be
|\<\psi_{00}|\phi_{00}\>|^2=\sqrt{P}
\ee
as long as the small deviation of $\sqrt{P}$ from $1$ is dominated by
the lowest order term in the effective interaction strength, as it is
the case in the example in {\bf Sect.\ref{leckm}}.\\
This means that no ``local'' errors are generated in first order, any deviation
occuring due to the factorization-approximation leads to entanglement
and therefore reduces purity. 
\section{Application to the Jaynes-Cummings-Model}
\label{leckm}
\subsection{The model}
The factorization-approximation will now be applied to the
Jaynes-Cummings-model. This choice is motivated by the fact that this model
can be solved exactly, hence it can be tested whether the approximation is
really applicable. Furthermore, both methods of treating this system, the
fully quantum mechanical and the semi-classical one, are well known, thus
the result of applying the factorization-approximation can easily be interpreted.\\
The Jaynes-Cummings-model \cite{BRU93} describes a spin in a magnetic field, interacting with some monochromatic electromagnetic field. A good example is a typical NMR experiment of any sort.\\
The model in rotating wave approximation is defined by the Hamiltonian:
\be
\hat{H}=Bg\mu\frac{1}{2}\hat{\sigma}_z+\hbar\omega(\hat{a}^{\dagger}\hat{a}+\frac{1}{2})+\hbar \gamma(\hat{a}^{\dagger}\hat{\sigma}^-+\hat{a}\hat{\sigma}^+),
\ee
where $B$ denotes the magnetic field, $g$ the gyromagnetic relation, $\mu$
Bohr's magneton, $\hat{\sigma}_z$ the operator of the z-component of the
spin, $\gamma$ the coupling constant, $\hat{\sigma}^+,\hat{\sigma}^-$ are
the creation and anihilation operators of the spin
system, $\omega$ is the freqency of the electromgnetic field and
$\hat{a}^{\dagger},\hat{a}$ are the creation and anihilation operators of the electromgnetic field.\\
The second term describes the monochromatic electromagnetic field, the
first term the spin in the magnetic field and the third term their mutual interaction.\\
\subsection{Application of the factorization-approximation}
Applying the scheme (\ref{fakt}) to this system yields:
\ba
&&i\hbar\frac{\partial}{\partial t}|\phi^S\>\\
&&=\left(Bg\mu\frac {1}{2}\hat{\sigma}_z+\hbar \gamma\left(\<\phi^L|\hat{a}^{\dagger}|\phi^L\>\hat{\sigma}^-+\<\phi^L|\hat{a}|\phi^L\>\hat{\sigma}^+\right)\right)|\phi^S\>\n\\
&&i\hbar\frac{\partial}{\partial t}|\phi^L\>\\
&&=\left(\hbar\omega\left(\hat{a}^{\dagger}\hat{a}+\frac {1}{2}\right)+\hbar \gamma\left(\<\phi^S|\hat{\sigma}^-|\phi^S\>\hat{a}^{\dagger}+\<\phi^S|\hat{\sigma}^+|\phi^S\>\hat{a}\right)\right)|\phi^L\>.\n
\ea
where $S$ indicates the spin system and $L$ the system of the
electromagnetic field.
The absolute values of the expectation values $\<\phi^S|\hat{\sigma}^-\phi^S\>, \<\phi^S|\hat{\sigma}^+\phi^S\>$ are always limited by:
\be
|\<\phi^S|\hat{\sigma}^-\phi^S\>|,|\<\phi^S|\hat{\sigma}^+\phi^S\>|\leq1.
\ee
The influence of the coupling terms on the evolution of the
electromagnetic field is thus negligible, if $\gamma\ll\omega$ and/or the system of the electromagnetic field is in a highly exited state. Thus, for example, a coherent state $|\alpha\>$ with a large parameter $\alpha$ is a valid solution to the equation controling the dynamics of the electromagnetic field.\\
Inserting this solution into the equation for the spin systems leads to:
\ba
&&i\hbar\frac{\partial}{\partial t}|\phi^S\>\\
&&=\left(Bg\mu\frac {1}{2}\hat{\sigma}_z+\hbar \gamma\left(\<\alpha|\hat{a}^+|\alpha\>\hat{\sigma}^-+\<\alpha|\hat{a}|\alpha\>\hat{\sigma}^+\right)\right)|\phi^S\>\n\\
&&=\left(Bg\mu\frac {1}{2}\hat{\sigma}_z+\hbar \gamma\sqrt{2}|\alpha|\left(\cos(\omega t)\hat{\sigma}_x+\sin(\omega t)\hat{\sigma}_y\right)\right)|\phi^S\>,\n
\ea 
This is exactly the semi-classical approach, which often produces exellent results, such as the correct Rabi-frequency etc.\\
It remains to be shown, that it was indeed justified to use the
factorization-approximation for some time $\tau$.\\
\subsection{Estimation of $P$}
If the spin and electromagnetic field are in resonance, an exact solution of the Jaynes-Cummings-model can be found:
\ba
|n^+\>:&=&\frac{1}{\sqrt{2}}(|0\>\otimes|n\>+|1\>\otimes|n-1\>\\
|n^-\>:&=&\frac{1}{\sqrt{2}}(|0\>\otimes|n\>-|1\>\otimes|n-1\>,\n
\ea
which satisfies the following eigenvalue equations:
\ba
\hat{H}|n^+\>&=&(\hbar\omega n+\hbar \gamma\sqrt{n})|n^+\>\\
\hat{H}|n^-\>&=&(\hbar\omega n-\hbar \gamma\sqrt{n})|n^-\>.\n
\ea
Considering the initial product state
\be
|\psi(0)\>=\left(\sum_nA_n|n\>\right)\otimes|0\>,
\ee 
we then find for the components of the Bloch vector of the spin system:
\ba
\label{blocomp}
\<\hat{\sigma}_x\otimes\hat{1}\>&=&\sum_n-\mbox{Im}\left(A_n^{\ast}A_{n+1}e^{-i\omega t}\right) \left( \sin \left( \gamma \left(\sqrt{n}+\sqrt{n+1}\right) t \right)\right.\n\\
&-&\left.\sin \left( \gamma \left( \sqrt{n}-\sqrt{n+1}\right) t \right) \right)\n\\
\<\hat{\sigma}_y\otimes\hat{1}\>&=&\sum_n\mbox{Re}\left(A_n^{\ast}A_{n+1}e^{-i\omega t}\right) \left( \sin \left( \gamma\left(\sqrt{n}+\sqrt{n+1}\right) t \right)\right.\n\\&-&\left.\sin \left( \gamma \left( \sqrt{n}-\sqrt{n+1}\right) t \right) \right)\n\\
\<\hat{\sigma}_z\otimes\hat{1}\>&=&\sum_n|A_n|^2\cos\left(2\sqrt{n}t\right).
\ea
From these components we find $P$ as:
\be
\label{rein}
\mbox{Tr}_S\lk\hat{\rho}^2_{S}\rk=P=\frac{1}{2}(1+\<\hat{\sigma}_x\>^2+\<\hat{\sigma}_y\>^2+\<\hat{\sigma}_z\>^2)
\ee
For the initial state we again choose a coherent state (for simplicity with
a real parameter $\alpha$), which means:
\be
A_n=e^{-\frac{1}{2}|\alpha|^2}\frac{\alpha^n}{\sqrt{n!}}.
\ee
Now we want to simplify the terms in (\ref{blocomp}). Consider:
\ba
\label{poiss}
A_n^{\ast}A_{n+1}&=&e^{-|\alpha|^2}\frac{\alpha^n}{\sqrt{n!}}\frac{\alpha^{n+1}}{\sqrt{\left(n+1\right)!}}\\
&=&\left(e^{-|\alpha|^2}\frac{\alpha^{2n}}{n!}\right)\left(\frac{\alpha}{\sqrt{\left(n+1\right)}}\right)\n.
\ea
The first factor describes a Poisson-distribution which is characterized by
the mean value $\alpha^2$  and the standard deviation $\alpha$. Since the
main weight of such a distribution is concentrated near its mean value
(98\% within 3 standard deviations), it is a reasonable approximation to
keep only contributions from this range. The second factor can be estimated
using its respective value at the boundaries of this range:
\be
\frac{\alpha}{\sqrt{\left(\alpha^2+\alpha+1\right)}}\leq\frac{\alpha}{\sqrt{\left(n_{rel}+1\right)}}\leq\frac{\alpha}{\sqrt{\left(\alpha^2-\alpha+1\right)}}
\ee
where $n_{rel}$ denotes the $n$'s from within this relevant range. If we now again only consider highly exited coherent states, that means the limit $\alpha\rightarrow\infty$, the upper as well as the lower bound converges against 1. Hence in this limit:
\be
\frac{\alpha}{\sqrt{\left(n_{rel}+1\right)}}\approx1.
\ee
In the same limit we also find:
\be
\sqrt{n}+\sqrt{\left(n+1\right)}\approx2\sqrt{n} \qquad \sqrt{n}-\sqrt{\left(n+1\right)}\approx0.
\ee
For a large enough $\alpha$ (\ref{blocomp}) therefore simplifies to:
\ba
\<\hat{\sigma}_x\otimes\hat{1}\>&=&\sin(\omega t)\sum_n|A_n|^2\sin \left(2\gamma\sqrt{n}t\right)\\
\<\hat{\sigma}_y\otimes\hat{1}\>&=&\cos(\omega t)\sum_n|A_n|^2\sin \left(2\gamma\sqrt{n}t\right)\n\\
\<\hat{\sigma}_z\otimes\hat{1}\>&=&-\sum_n|A_n|^2\cos\left(2\gamma\sqrt{n}t\right).\n
\ea
In this approximation the factors depending on $\omega t$ describe the
Larmor-precession, while the sum terms give rise to the Rabi-precession.\\
Expanding the argument of the Rabi-precession arround the mean value of the
Poisson-distribution yields:
\ba
&&2\gamma\sqrt{n}t\approx\\ 
&&\gamma t\left(2\alpha+\frac {1}{\alpha}(n-\alpha^2)-\frac {1}{2\alpha^3}(n-\alpha^2)^2+O^3(n-\alpha^2)\right)\n
\ea
Since we are only considering contributions from the relevant region, that
is contributions with $|n-\alpha^2|\leq\alpha$ in the limit
$\alpha\rightarrow\infty$, we can neglect any term beyond second-order and get:
\be
\label{oszi}
2\gamma\sqrt{n}t\approx\gamma t \left(2\alpha+\frac{(n-\alpha^2)}{\alpha}\right).
\ee
Using this approximation and some addition theorems we get on the right hand
side of (\ref{rein}) fast oscillating terms with a frequency given by the
first term in (\ref{oszi}) and slow oscillating terms with a frequency
given by the second term in (\ref{oszi}). Since the slow oscillating terms
from the relevant region oscillate with a frequency of the order of
$\gamma$ regardless of the actual $\alpha$, we expand those in time to
second order, which will be a good approximation as long as $\gamma t$
remains small: 
\be
P=1-\gamma^2 t^2
\ee
This means that in the limit of high intensities of the electromagnetic
field, the ``coherence time'' is independent of the intensity itself
and only depends on the strength of the coupling. If that is weak enough,
the factorization approximation will be valid for a considerable time.

\section{Summary and conclusions}
We have shown that bi-partite systems with mutual interactions, that
remain unentangled  forever, no matter which initial product state was
chosen, cannot exist.\\ 
One might argue that this is a rather weak statement: In large systems with many degrees of
freedom there might be a large number of initial states, starting from which
the subsystems would remain entanglementfree, even though there must also be some
initial states that lead to entanglement.\\ 
But the same could be argued, for
example, for a classical Hamilton-system, which has no radial symmetry, say: Although
angular momentum is not conserved in such a situation in general, there
might be special trajectories for which it is conserved. But one probably
would not claim that these trajectories are of importance considering
the space of all possible trajectories, especially in systems with
many degrees of freedom.\\
The same holds true for the entanglement conservation: there might be
special initial states that prevent the systems from getting entangled, but
it can be shown that their relative weight decreases with the system size.\\
Thus it is, strictly speaking, unjustified to describe a particle in a box,
which is part of an interacting quantum system, by a wave-function.\\
It is, nevertheless, a potentially very good approximation to describe
interacting systems by pure individual wavefunctions, if the entanglement
remains small during the time of observation. The underlying equation can
then be based on the factorization approximation; this equation has been derived, together with a criterion for its validity.\\
For a spin interacting with a monochromatic electromagnetic field, we have shown that the criterion is fulfilled for high enough field intensities
and for a time that only depends inversely on the coupling
strength. Applying the factorization approximation then transforms the full
quantum mechanical Jaynes-Cummings-model into a semi-classical one, in
which the field acts as an external potential.

\acknowledgements
We thank C.~Granzow, A.~Otte, I.~Kim, F.~Tonner and M.~Stollsteimer for fruitful discussions.


\begin{thebibliography}{9}
\addcontentsline{toc}{chapter}{Literaturverzeichnis}

\bibitem{EIN35} Einstein,~A., Podolsky,~B., Rosen,~N. Can Quantum Mechanical Description of Physical Reality be Considered Complete ?
{\em Phys.~Rev.} {\bf 47}, 777 (1935).

\bibitem{OTT98} Otte,~A. Physik und Quanteninformation - Darstellung, Symmetrie und Dynamik (Diplomarbeit, Stuttgart 1998)

\bibitem{DEU85} Deutsch,~D. Quantum theory, the Church-Turing principle and the universal quantum computer
{\em Phys.~Let.} {\bf 79}, 325-328 (1997).

\bibitem{SHO94} Shor,~P.~W in 
{\em Proc. of the Symp.on the Foundations of Computerscience, Los Alamitos,
  California} 124-134 (IEEE Computer Society Press, New York, 1994).

\bibitem{GRO97} Grover,~L. Quantum Mechanics Helps in  Searching for a
  Needle in a Haystack
{\em Proc. R. Soc. Lond. A} {\bf 400}, 97-117 (1985).

\bibitem{CHU98} Chuang,~I., Gershenfield,~N., Kubinec,~M. Experimental
  Implementation of Fast Quantum Searching
{\em Phys.~Let.} {\bf 80}, 3408 (1998).

\bibitem{TUR98} Turchette,~Q. et al. Deterministic Entanglement of Two
  Trapped Ions
{\em Phys.~Let.} {\bf 81}, 3631 (1998).

\bibitem{MAH98} Mahler,~G., Weberruss,~V. Quantum Networks (Springer, Berlin, 1998)

\bibitem{VED97} Vedral,~V., Plenio,~M., Rippin,~M., Knight,~P. Quanifying Entanglement
{\em Phys.~Let.} {\bf 78}, 2275 (1997).

\bibitem{SCH95} Schlienz,~J., Mahler,~G. Description of Entanglement
{\em Phys.~Rev.~A} {\bf 52}, 4396-4404 (1995).

\bibitem{ESP90} D'Espagnat,~B. Towards a Seperable Empirical Reality ? 
{\em Foundations of Physics} {\bf 20}, 1147 (1990).

\bibitem{HAR28} Hartree,~D.
{\em Proc. Cambridge Phil. Soc.} {\bf 24}, 89 (1928). 

\bibitem{BRU93} Shore,~B.~W., Knight,~P. The Jaynes-Cummings-Model
{\em Journal of modern Optics} {\bf 40} (1993)

\bibitem{LUI98} De la Pena,~L., Santos,~E. Pertubation of the evolution of a quantum system induced by its environment {\em Phys.~Let.~A} {\bf 259} 83-90 (1999)

\bibitem{KAN98} Kane,~E. A silicon based nuclear spin quantum computer
{\em Nature} {\bf 393}, 133 (1998).

\bibitem{JIA98} Jian-Wei Pan, Bouwmeester,~D., Zeillinger,~A. Entangling
  Photons That Never Interacted.
{\em Phys.~Let.} {\bf 80}, 3819 (1998). 

\end{thebibliography}
\end{document}